\pdfoutput=1
\documentclass[pre,aps,preprint,superscriptaddress,amsmath,amssymb]{revtex4-2}
\usepackage{amsthm}
\usepackage{mathtools}
\usepackage{graphicx}
\usepackage{booktabs}
\usepackage[colorlinks=true,allcolors=blue]{hyperref}

\theoremstyle{definition}

\newcommand{\ii}{\mathrm{i}}

\begin{document}

\title{Integrable discretization of the Maxwell--Bloch equations}

\author{I.~Kuk}
\author{I.~Gabitov}
\email{gabitov@arizona.edu}
\affiliation{Department of Mathematics, University of Arizona, Tucson, Arizona 85721, USA}

\begin{abstract}
Numerical schemes for the Maxwell--Bloch equations keep the length of the Bloch vector but not
the integrable structure, so that solitons are only approximate solutions. We write an integrable
discretization, in both time and space, of these equations for a complex field in a resonant
two-level medium in the physical variables and use it as a numerical scheme. The lattice follows from a
discrete Lax pair with the analytic structure of the continuous one; an auxiliary field variable
is fixed by an isospectral normalization, and the resulting scheme is explicit, a $2\times2$
similarity transformation of the density matrix per grid node. It conserves the length of every
Bloch vector and satisfies an exact energy balance between field and medium. Viewed as an
exchange interaction between a field spin and the Bloch vector, the lattice coincides with a
known fully discrete integrable system, which we use to obtain closed-form solitons and
breathers; they are reflectionless potentials with energy $16\sum_j\mathrm{Im}\,\lambda_j$, where
$\lambda_j$ are the eigenvalues, and converge to the
McCall--Hahn soliton and Lamb's breather. Inhomogeneous broadening is included through ordered
sublayers that keep integrability. On a finite time window the scattering data evolve by a
closed map, with the final state of the medium given by the data at one point of the spectral
plane. On the infinite line, if the medium returns to its ground state behind the pulse (which
we show on the first row and observe numerically on later rows), the transmission coefficient is
conserved and the reflection coefficient evolves linearly, with the multiplier of the linearized
lattice. As a numerical method the lattice is explicit and second order, with a cost per accuracy
comparable to that of Cayley and exponential-midpoint schemes, and an area-preserving input map
with a calibrated medium coupling makes it exact for a chosen soliton.
\end{abstract}

\maketitle

\section{Introduction}
\label{sec:intro}

A short optical pulse in a resonant medium of two-level atoms can pass without loss as a
self-induced transparency (SIT) soliton~\cite{McCallHahn1969}; the same medium produces photon
echoes and, when inverted, superfluorescence~\cite{AllenEberly1975,GabitovZakharovMikhailov1984}.
In the absence of relaxation these phenomena are described by the Maxwell--Bloch (MB)
equations, which are completely
integrable~\cite{Lamb1971,AblowitzKaupNewell1974,GabitovZakharovMikhailov1985}. The inverse
scattering transform gives their solitons and breathers, the area theorem and the asymptotic
reshaping of arbitrary pulses, and it remains an active subject, with recent work on nonzero
backgrounds and inhomogeneous
broadening~\cite{BiondiniGabitovKovacicLi2019,AbeyaBiondiniKovacicPrinari2024}.

In practice the MB equations are solved numerically, for example by the Crank--Nicolson
predictor--corrector scheme of Ziolkowski \textit{et al.}~\cite{Ziolkowski1995} for the full-wave
equations without the rotating-wave approximation, or, for the envelope equations, by
time discretizations that decouple the field and the medium~\cite{Bidegaray2003}. The length of
the Bloch vector is preserved exactly when the vector is rotated by the implicit midpoint
(Cayley) rule solved to convergence~\cite{HairerLubichWanner2006,IserlesMuntheKaasNorsettZanna2000}
or by exact rotations within a splitting
scheme~\cite{McLachlanQuispel2002,BidegarayBourgeadeReignier2001}.
These schemes do not keep the integrable structure of the equations: a soliton is not an exact
solution of the scheme, its energy and area are conserved only approximately, and its position
drifts slowly over long distances. For the nonlinear Schr\"odinger equation, nonintegrable
discretizations can even produce numerically induced chaos, which an integrable discretization
avoids~\cite{HerbstAblowitz1989}; integrable discretizations are not free of numerical
difficulties either, as shown for a doubly discrete sine-Gordon scheme near homoclinic
manifolds~\cite{AblowitzHerbstSchober1996,AblowitzHerbstSchober1997}.

Integrable discretizations keep this structure: the lattice equation has its own Lax pair, exact
soliton solutions and conservation
laws~\cite{AblowitzLadik1975,AblowitzLadik1976,Suris2003,BobenkoSuris2008}; for equations on
quad-graphs the Lax pair follows from the consistency of the equation on a
cube~\cite{BobenkoSuris2002,AdlerBobenkoSuris2003}. For the MB
equations, integrable lattices with continuous time were found by Bogoliubov
\textit{et al.} and Rybin \textit{et al.}~\cite{Bogoliubov1995,Rybin2001}, and partial
discretizations were studied by Saksida~\cite{Saksida2006}; for real fields, the fully discrete
problem reduces to Hirota's discrete sine-Gordon
equation~\cite{Hirota1977c,Orfanidis1978,BobenkoPinkall1996}. To our knowledge, an integrable
discretization of the complex MB system in both space and time has not been written in the
physical variables nor used as a numerical scheme.

In this paper we write such a discretization in the physical variables $E$, $\rho$, $N$ and study
it both as a model of two interacting classical spins, the field and the Bloch vector, and as a
numerical method. The lattice follows from a discrete Lax pair with the analytic structure and
symmetry of the continuous one (Sec.~\ref{sec:discretization}); it is explicit, conserves the
length of every Bloch vector and obeys an exact energy balance between field and medium. As a
lattice it is not new: it coincides with a fully discrete system of Tsuchida~\cite{Tsuchida2015} related to the
Yang--Baxter maps of Goncharenko and Veselov~\cite{GoncharenkoVeselov2004} (for Lax matrices of
Yang--Baxter maps see~\cite{SurisVeselov2003}). What is new is its reading and use as a
Maxwell--Bloch scheme. Using the connection, we obtain the soliton and the breather in closed form
(Sec.~\ref{sec:soliton}), compare the scheme with standard methods (Sec.~\ref{sec:numerics}) and
extend it to inhomogeneously broadened media by ordered sublayers (Sec.~\ref{sec:broadening}).
For the scattering data (Sec.~\ref{sec:scattering}) we obtain lattice counterparts of the equations of the continuous
theory~\cite{GabitovZakharovMikhailov1985}: a closed map on a finite time window
and, on the infinite line, provided the medium returns to its ground state behind the pulse, a
linear evolution of the reflection coefficient with the multiplier of the linearized lattice; for
the real reduction this linear law was found by Boiti \textit{et al.}~\cite{BoitiPempinelliPrinariSpire2002}.
Unlike the continuous theory, the transmission coefficient is then conserved; how the two
pictures connect in the continuum limit remains open. As a numerical scheme the lattice is
explicit, second order and needs no iteration; its cost per accuracy is comparable to that of
standard structure-preserving schemes, and with a calibrated input it reproduces a chosen
soliton exactly.

Section~\ref{sec:prelim} recalls the model, Sec.~\ref{sec:discretization} derives the lattice
and its properties, Sec.~\ref{sec:soliton} gives exact solutions, Sec.~\ref{sec:numerics}
compares the scheme with standard methods, Sec.~\ref{sec:broadening} treats inhomogeneous
broadening, Sec.~\ref{sec:scattering} derives the evolution of the scattering data, and
Sec.~\ref{sec:discussion} summarizes the results and open problems. Derivations
are collected in the Appendixes.

\section{Model}
\label{sec:prelim}

We consider the propagation of a coherent optical pulse through a resonant
medium of two-level atoms in the absence of relaxation. In the frame moving
with the pulse and in dimensionless units, the envelope $E(x,t)$ of the
electric field and the density matrix
\[
  \hat\rho=\begin{pmatrix}N&\rho\\ \rho^*&-N\end{pmatrix}
\]
of the atoms, with polarization $\rho$ and population inversion $N$, satisfy
the Maxwell--Bloch (MB) equations
\cite{McCallHahn1969,Lamb1971,AllenEberly1975,AblowitzKaupNewell1974,GabitovZakharovMikhailov1985}
\begin{equation}
  E_x=\rho,\qquad \rho_t+2\ii\lambda_0\rho=NE,\qquad
  N_t=-\tfrac12\bigl(E^*\rho+E\rho^*\bigr).
  \label{eq:MB-full}
\end{equation}
Here $x$ is the propagation distance, $t$ is the retarded time, and
$\lambda_0$ is the detuning of the atomic transition from the carrier
frequency. For a single (sharp) spectral line, writing
$\rho=\tilde\rho\,e^{-2\ii\lambda_0t}$, $E=\tilde E\,e^{-2\ii\lambda_0t}$ removes
$\lambda_0$ from the equations for $\tilde E,\tilde\rho$, and we set $\lambda_0=0$ below:
\begin{equation}
  E_x=\rho,\qquad \rho_t=NE,\qquad N_t=-\tfrac12\bigl(E^*\rho+E\rho^*\bigr);
  \label{eq:MB}
\end{equation}
the inhomogeneously broadened medium is treated in Sec.~\ref{sec:broadening}.
Equations~\eqref{eq:MB-full} conserve the length of the Bloch vector, and we
normalize it as $N^2+|\rho|^2=1$. We consider the signaling problem: the
input field $E(0,t)$ is given for all $t$, and before the pulse arrives the
medium is in a known state, $N=N_-$, $\rho=0$ (for an absorbing medium in the
ground state $N_-=-1$).

Equations~\eqref{eq:MB-full} are the compatibility condition of the
overdetermined linear system
\cite{AblowitzKaupNewell1974,GabitovZakharovMikhailov1985}
\begin{equation}
  \Psi_t=U\Psi,\qquad \Psi_x=V\Psi,
  \label{eq:lax}
\end{equation}
\begin{equation}
  U=-\ii\lambda\sigma_3-\tfrac12u,\quad
  u=\begin{pmatrix}0&E\\-E^*&0\end{pmatrix},\qquad
  V=-\frac{\hat\rho}{4\ii(\lambda-\lambda_0)},
  \label{eq:UVcont}
\end{equation}
for a $2\times2$ matrix wave function $\Psi(x,t,\lambda)$; the first equation is the
Zakharov--Shabat spectral problem~\cite{ZakharovShabat1972} in the form of
Ablowitz, Kaup, Newell and Segur~\cite{AblowitzKaupNewellSegur1974}. Equality of the
mixed derivatives, $\Psi_{tx}=\Psi_{xt}$, gives $U_x-V_t+[U,V]=0$, which is
equivalent to~\eqref{eq:MB-full}. The spectral parameter $\lambda$ enters $U$
linearly and $V$ through a simple pole at the line center $\lambda=\lambda_0$.
Both matrices obey the symmetry
$\sigma_2U(\lambda)\sigma_2=U^*(\lambda^*)$,
$\sigma_2V(\lambda)\sigma_2=V^*(\lambda^*)$, which encodes the reality of $N$
and the conjugate pairs $(E,E^*)$ and $(\rho,\rho^*)$. In what follows we
discretize~\eqref{eq:lax} on a rectangular lattice in $(t,x)$ so that this
analytic structure and this symmetry are kept.

\section{The integrable lattice}
\label{sec:discretization}
In this section we derive the lattice from a discrete Lax pair, fix its auxiliary variable by an isospectral normalization, write the explicit step and its conservation laws, and relate the lattice to known integrable systems.

\subsection{Discrete Lax pair and lattice equations}
\label{sec:laxpair}
\begin{figure}[t]
\includegraphics[width=\columnwidth]{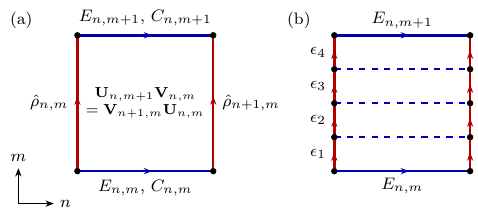}
\caption{(a) Elementary cell of the lattice. The field variables $E,C$ sit on the time edges and
the density matrix $\hat\rho$ on the space edges; the explicit step~\eqref{eq:qstep} computes the
upper and right edges from the lower and left ones. (b) For inhomogeneous broadening a space step
is split into ordered sublayers, each carrying one detuning class $\epsilon_j$
(Sec.~\ref{sec:sublayers}); the field passes the sublayers in turn.}
\label{fig:lattice}
\end{figure}

We place the wave function $\Psi_{n,m}$ at the nodes $(t_n,x_m)=(n\Delta t,m\Delta x)$
of a rectangular lattice and replace~\eqref{eq:lax} by the shifts
\begin{equation}
  \Psi_{n+1,m}=\mathbf U_{n,m}\Psi_{n,m},\qquad \Psi_{n,m+1}=\mathbf V_{n,m}\Psi_{n,m},
  \label{eq:dlax}
\end{equation}
whose compatibility condition is
\begin{equation}
  \mathbf U_{n,m+1}\mathbf V_{n,m}=\mathbf V_{n+1,m}\mathbf U_{n,m}.
  \label{eq:compat}
\end{equation}
The field lives on the time edges $(n,m)\to(n+1,m)$ and the medium on the space
edges $(n,m)\to(n,m+1)$ [Fig.~\ref{fig:lattice}(a)].

 The naive choice $\mathbf U=I+\Delta t\,U$,
$\mathbf V=I+\Delta x\,V$ fails: \eqref{eq:compat} must hold for all $\lambda$,
which gives seven independent real conditions for the five real unknowns $E_{n,m+1}$,
$\rho_{n+1,m}$, $N_{n+1,m}$, and for generic data these conditions have no solution.

We therefore look for a pair with the same analytic structure as
\eqref{eq:UVcont} (a linear dependence on $\lambda$ for $\mathbf U$, a simple pole
at $\lambda=0$ for $\mathbf V$) and the same $\sigma_2$-symmetry, and we require
that the zeros of $\det\mathbf U_{n,m}$ do not depend on $n$ and $m$. After a gauge
transformation and the absorption of the lattice steps into the variables
($\lambda\to\Delta t\,\lambda$, $E\to\Delta t\,E$, $\hat\rho\to\Delta t\Delta x\,\hat\rho$;
Appendix~\ref{app:derivation}), the pair becomes
\begin{equation}
  \mathbf U_{n,m}=-\ii\lambda\sigma_3-\tfrac12u_{n,m}+(1+C_{n,m})\,I,\qquad
  \mathbf V_{n,m}=I-\frac{\hat\rho_{n,m}}{4\ii\lambda}.
  \label{eq:Unm}
\end{equation}
The only new ingredient compared with the naive pair is the real scalar $C_{n,m}$ on
the time edges. It has no counterpart in~\eqref{eq:MB-full}; we show below that it
is fixed by the field and vanishes in the continuum limit as $O(\Delta t^2)$. Unless stated otherwise, symbols are in these lattice units in Secs.~\ref{sec:discretization}, \ref{sec:soliton} and~\ref{sec:scattering} and in physical units in Sec.~\ref{sec:numerics}; the subscripts ``ph'' and ``lat'' mark the units where both appear.

\label{sec:latticeeq}
Substituting~\eqref{eq:Unm} into~\eqref{eq:compat} (the terms linear in $\lambda$ cancel
identically) and equating the terms constant in $\lambda$ and the residue at $\lambda=0$, we
obtain
\begin{subequations}\label{eq:lattice}
\begin{align}
  E_{n,m+1}&=E_{n,m}+\tfrac12(\rho_{n,m}+\rho_{n+1,m}),\label{eq:lattice1}\\
  C_{n,m+1}&=C_{n,m}-\tfrac14(N_{n,m}-N_{n+1,m}),\label{eq:lattice2}\\
  \rho_{n+1,m}&=\rho_{n,m}+\tfrac12\bigl(E_{n,m+1}N_{n,m}+E_{n,m}N_{n+1,m}\bigr)
     +\rho_{n,m}C_{n,m+1}-\rho_{n+1,m}C_{n,m},\label{eq:lattice3}\\
  N_{n+1,m}&=N_{n,m}-\tfrac12\bigl(E_{n,m+1}\rho^*_{n,m}+E^*_{n,m}\rho_{n+1,m}\bigr)
     +N_{n,m}C_{n,m+1}-N_{n+1,m}C_{n,m}.\label{eq:lattice4}
\end{align}
\end{subequations}
For given $(E,C)$ on the time edge and $\hat\rho$ on the space edge leaving the node
$(n,m)$, Eqs.~\eqref{eq:lattice} are the unique solution of~\eqref{eq:compat} for the
four unknowns $E_{n,m+1}$, $C_{n,m+1}$, $\rho_{n+1,m}$, $N_{n+1,m}$.
Equation~\eqref{eq:lattice1} is the midpoint rule for $E_x=\rho$ in $x$, with $\rho$ averaged
over the two adjacent times;
Eqs.~\eqref{eq:lattice3} and~\eqref{eq:lattice4} are a discrete Bloch rotation of
the atoms on the space edge driven by the field on the two adjacent time edges;
Eq.~\eqref{eq:lattice2} transports the auxiliary variable $C$. The
$\sigma_2$-symmetry of~\eqref{eq:Unm} is preserved by~\eqref{eq:compat}, so $C$
and $N$ stay real.

\subsection{Normalization and explicit step}
\label{sec:normalization}
The determinants of the Lax matrices are
\begin{equation}
  \det\mathbf U_{n,m}=\lambda^2+\kappa_{n,m}^2,\quad \kappa^2=(1+C)^2+\tfrac14|E|^2,
  \qquad
  \det\mathbf V_{n,m}=1+\frac{r_{n,m}^2}{16\lambda^2},\quad r^2=N^2+|\rho|^2 .
  \label{eq:dets}
\end{equation}
Taking the determinant of~\eqref{eq:compat}, we find that
$\kappa_{n,m+1}=\kappa_{n,m}$ whenever $r_{n+1,m}=r_{n,m}$; the latter is shown in
Sec.~\ref{sec:invariants}. Each time edge thus keeps its value of $\kappa$ as the
pulse propagates. Our requirement that the zeros $\lambda=\pm\ii\kappa$ of
$\det\mathbf U$ be the same on all edges makes $\kappa$ a global constant, which a
rescaling of $\lambda$ and $\Delta t$ sets to one:
\begin{equation}
  1+C_{n,m}=\sqrt{1-\tfrac14|E_{n,m}|^2}.
  \label{eq:normalization}
\end{equation}
It suffices to impose~\eqref{eq:normalization} on the input row $m=0$;
Eqs.~\eqref{eq:lattice} then keep it everywhere. Condition~\eqref{eq:normalization}
requires $|E|<2$ in lattice units, i.e., $|E|\Delta t<2$ in physical units, and gives
$C=-\tfrac18|E|^2\Delta t^2+O(\Delta t^4)$, so that $C\equiv0$ is excluded by the
normalization unless $E\equiv0$. The positive root is kept from $m$ to $m+1$ on every time edge
as long as $|E_{n,m}|^2<4-r^2$, since $|C_{n,m+1}-C_{n,m}|\le r/2$ by~\eqref{eq:lattice2}; this
holds for small steps. With~\eqref{eq:normalization}
the free lattice spectral problem has the Jost solutions $(1\mp\ii\lambda)^n$, and
its continuous spectrum is the real axis.

\label{sec:explicit}
Equations~\eqref{eq:lattice3} and~\eqref{eq:lattice4} are equivalent to a similarity
transformation of the density matrix,
\begin{equation}
  \hat\rho_{n+1,m}=G_{n,m}\,\hat\rho_{n,m}\,G_{n,m}^{-1},\qquad
  G_{n,m}=(1+C_{n,m})\,I-\tfrac12u_{n,m}-\tfrac14\sigma_3\hat\rho_{n,m},
  \label{eq:qstep}
\end{equation}
where $G$ depends only on the data at the node $(n,m)$. The step is therefore
explicit: one computes $G$, inverts a $2\times2$ matrix, updates $\hat\rho$, and then
obtains $E_{n,m+1}$ and $C_{n,m+1}$ from~\eqref{eq:lattice1} and~\eqref{eq:lattice2}.
It is well defined while
\begin{equation}
  \det G=\bigl(1+C-\tfrac14N\bigr)^2+\bigl|\tfrac12E+\tfrac14\rho\bigr|^2\neq0 .
  \label{eq:detq}
\end{equation}
With~\eqref{eq:normalization} and $r<2\sqrt2$ the right-hand side is positive for
every field and atomic state: if $|E|\le r/2$, then
$1+C\ge\sqrt{1-r^2/16}>r/4\ge N/4$; otherwise the second term is positive. In
physical units $r=\Delta t\,\Delta x$, so this condition is far from restrictive.

\subsection{Conservation laws}
\label{sec:invariants}
Since~\eqref{eq:qstep} is a similarity transformation, the eigenvalues $\pm r$ of
$\hat\rho$ are conserved along $t$: every atom keeps the length of its Bloch vector
exactly, $N_{n+1,m}^2+|\rho_{n+1,m}|^2=N_{n,m}^2+|\rho_{n,m}|^2$, for any step size.
The field has the analogous invariant $\kappa$ of Sec.~\ref{sec:normalization},
conserved along $x$. Both invariants are determinants of the Lax
matrices~\eqref{eq:dets}; their meaning as spin lengths is explained in
Sec.~\ref{sec:structure}.

\label{sec:energy}
Summing~\eqref{eq:lattice2} over a time window $0\le n<n_{\max}$ gives
\begin{equation}
  W_{m+1}-W_m=-2\bigl(N_{n_{\max},m}-N_{0,m}\bigr),\qquad
  W_m=-8\sum_nC_{n,m}=8\sum_n\Bigl(1-\sqrt{1-\tfrac14|E_{n,m}|^2}\Bigr).
  \label{eq:energy}
\end{equation}
Here $W=\sum_n|E_{n,m}|^2\bigl[1+O(|E|^2)\bigr]$ is the lattice energy of the pulse,
and~\eqref{eq:energy} is the exact lattice counterpart of
$\partial_x\!\int|E|^2dt=-2\,[N(t{=}{+}\infty)-N(t{=}{-}\infty)]$: the field loses
exactly the energy the medium gains. The naive quantity $\sum_n|E_{n,m}|^2$ is
conserved only up to $O(\Delta t^2)$.

\subsection{Continuum limit and relation to known systems}
\label{sec:convergence}
In physical variables ($E_{\rm lat}=\Delta t\,E$,
$\hat\rho_{\rm lat}=\Delta t\Delta x\,\hat\rho$), Eq.~\eqref{eq:lattice1} is the
midpoint rule $E(x+\Delta x)=E(x)+\tfrac{\Delta x}{2}[\rho(t_n)+\rho(t_{n+1})]$
for $E_x=\rho$, with $\rho$ taken at $x_{m+1/2}$ and averaged over two times, and
\[
  G=I-\tfrac{\Delta t}{2}\,u+O(\Delta t^2,\Delta t\,\Delta x),\qquad
  \hat\rho_{n+1,m}=\hat\rho_{n,m}+\Delta t\,\bigl[-\tfrac12u,\hat\rho_{n,m}\bigr]+\dots,
\]
which is the Bloch equation $\hat\rho_t=[-\tfrac12u,\hat\rho]$ contained
in~\eqref{eq:MB}. The lattice is thus a consistent discretization of~\eqref{eq:MB}
with the field on the time edges (at $t_{n+1/2}$) and the medium on the space edges
(at $x_{m+1/2}$). This consistency argument is formal; second-order convergence is
established numerically for the soliton and for generic pulses in Sec.~\ref{sec:numerics} and
for the breather in Sec.~\ref{sec:breather}.

\label{sec:structure}
Multiplying the Lax matrices by $\ii\sigma_3$ and by $\lambda$, respectively, we obtain
\begin{equation}
  \ii\sigma_3\mathbf U=\lambda I+\ii S_E,\quad
  S_E=\sigma_3\bigl[(1+C)I-\tfrac12u\bigr],\qquad
  \lambda\mathbf V=\lambda I+\ii S_B,\quad S_B=\tfrac14\hat\rho,
  \label{eq:spins}
\end{equation}
where $S_E$ and $S_B$ are traceless Hermitian matrices with $S_E^2=\kappa^2I$ and
$S_B^2=(r/4)^2I$. Both factors are rational Lax operators of the classical
Heisenberg (XXX) type~\cite{FaddeevTakhtajan1987}, and the compatibility
condition~\eqref{eq:compat}, with the updated Bloch spin $S_B''$ written as $\sigma_3S_B''\sigma_3$ (the
staggered gauge below), is the refactorization of a field spin $S_E$ on the time edges and a
Bloch spin $S_B$ on the space edges: each lattice step is an exact exchange interaction between
the two spins. Integrable space-time circuits of classical spin and matrix fields built from such
two-body maps were studied by Krajnik, Ilievski and Prosen~\cite{KrajnikIlievskiProsen2020}.
After the staggered gauge $\Psi_{n,m}\to\sigma_3^n\Psi_{n,m}$, which reverses
the signs of $E$ and $\rho$ on odd $n$, the lattice coincides with the scalar
Hermitian case of the fully discrete system of Tsuchida~\cite{Tsuchida2015}, with
the parameters $\alpha_{\rm T}=\ii\kappa$ and $a_{\rm T}=\ii r/4$ and the stereographic coordinates
$u_{\rm T}=-E/[2(\kappa-1-C)]$ and $q_{\rm T}=\rho/(r-N)$ (Appendix~\ref{app:tsuchida}); that
system generalizes the refactorization problem of Goncharenko and
Veselov~\cite{GoncharenkoVeselov2004}, and reduced forms of its refactorization and Yang--Baxter
map appeared earlier~\cite{TerngUhlenbeck2000,KouloukasPapageorgiou2011}. Tsuchida also identifies
the first negative flow of the nonlinear Schr\"odinger hierarchy with the SIT (MB) equations and
gives its integrable discretization in the propagation variable [Eq.~(2.21) of
Ref.~\cite{Tsuchida2015}]. For real fields, with
$1+C=\cos\alpha$, $E/2=\sin\alpha$, $N=r\cos\Theta$ and $\rho=r\sin\Theta$, a constant
gauge transformation maps~\eqref{eq:Unm} onto Orfanidis' Lax pair of Hirota's
discrete sine-Gordon equation~\cite{Hirota1977c,Orfanidis1978,BobenkoPinkall1996},
with Orfanidis' parameter $\gamma_{\rm O}=r/4$ and the spectral parameter $k_{\rm O}=-\lambda$.

Integrable MB lattices with continuous time were constructed
in Refs.~\cite{Bogoliubov1995,Rybin2001}; here both variables are discrete. To our
knowledge, the fully discrete lattice has not been interpreted as a discretization
of the MB equations, with the field and the Bloch vector as two interacting spins.
We use the known structure as a tool: it gives the exact solutions of
Sec.~\ref{sec:soliton} and the broadened lattice of Sec.~\ref{sec:broadening}.

\section{Exact solutions}
\label{sec:soliton}
Exact solutions of the lattice follow from the dressing method; we give the soliton and the breather in closed form and their continuum limits.

\subsection{Dressing on the lattice}
\label{sec:dressing}
Exact solutions of~\eqref{eq:lattice} follow from the Darboux (dressing)
method~\cite{MatveevSalle1991} applied to the discrete Lax pair~\eqref{eq:dlax}. We take
the medium at rest, $E=C=\rho=0$, $N=N_0$ on all edges, so that
$\hat\rho_0=N_0\sigma_3$, $\mathbf U_0=I-\ii\lambda\sigma_3$ and
$\mathbf V_0=I-N_0\sigma_3/(4\ii\lambda)$. The free wave function is
\begin{equation}
  \Psi^{(0)}_{n,m}(\lambda)=\begin{pmatrix}(1-\ii\lambda)^nc^m&0\\0&(1+\ii\lambda)^nd^m\end{pmatrix},
  \qquad c,d=1\mp\frac{N_0}{4\ii\lambda}.
  \label{eq:Psi0}
\end{equation}
We choose a point $\lambda_1=\xi+\ii\eta$ with $\eta>0$ and a constant $\gamma\neq0$, and set
$\psi_{n,m}=\Psi^{(0)}_{n,m}(\lambda_1)(1,\gamma)^T\propto(1,z_{n,m})^T$, where
\begin{equation}
  z_{n,m}=\gamma\,q^np^m,\qquad q=\frac{1+\ii\lambda_1}{1-\ii\lambda_1},\qquad
  p=\frac{1+N_0/(4\ii\lambda_1)}{1-N_0/(4\ii\lambda_1)}.
  \label{eq:zqp}
\end{equation}
With the orthogonal projector $\Pi_{n,m}=\psi\psi^\dagger/(\psi^\dagger\psi)$ we define
\begin{equation}
  \mathcal D_{n,m}(\lambda)=I+\frac{\bar\lambda_1-\lambda_1}{\lambda-\bar\lambda_1}\,\Pi_{n,m},
  \qquad \det\mathcal D=\frac{\lambda-\lambda_1}{\lambda-\bar\lambda_1}.
  \label{eq:dressing}
\end{equation}
The function $\Psi=\mathcal D\Psi^{(0)}$ satisfies~\eqref{eq:dlax} with
\[
  \tilde{\mathbf U}_{n,m}=\mathcal D_{n+1,m}\mathbf U_0\mathcal D_{n,m}^{-1},\qquad
  \tilde{\mathbf V}_{n,m}=\mathcal D_{n,m+1}\mathbf V_0\mathcal D_{n,m}^{-1}.
\]
The poles at
$\lambda_1$ and $\bar\lambda_1$ cancel because $\psi$ is transported by
$\mathbf U_0(\lambda_1)$ and $\mathbf V_0(\lambda_1)$ and $\psi^\perp$ by the conjugate
matrices. Hence $\tilde{\mathbf U}$ is linear in $\lambda$, $\tilde{\mathbf V}$ has a single
pole at $\lambda=0$, and both have exactly the form~\eqref{eq:Unm}. The new fields are read
off as
\begin{equation}
  E=-2\bigl(\tilde{\mathbf U}\bigr)_{12}\big|_{\lambda^0},\qquad
  1+C=\bigl(\tilde{\mathbf U}\bigr)_{11}\big|_{\lambda^0},\qquad
  \hat\rho_{n,m}=\mathcal D_{n,m+1}(0)\,\hat\rho_0\,\mathcal D_{n,m}(0)^{-1}.
  \label{eq:readoff}
\end{equation}
Since $\det\mathcal D$ does not depend on $n$ and $m$, $\det\tilde{\mathbf U}=\det\mathbf U_0=\lambda^2+1$
and $\det\tilde{\mathbf V}=\det\mathbf V_0$: the dressed solution satisfies
$(1+C)^2+\tfrac14|E|^2=1$ automatically and has $r=|N_0|$. For small lattice parameters,
$1+C>0$, so~\eqref{eq:normalization} holds with the positive root; for one point $\lambda_1$
this is the case when $8\eta^2<(|1-\ii\lambda_1|+|1+\ii\lambda_1|)^2$, i.e., $\eta<1/\sqrt2$
for $\xi=0$. For larger parameters $C$ is the value given by the dressing. In particular,
$C\neq0$ wherever $E\neq0$; no nontrivial solution of the lattice has $C\equiv0$.
Iterating the construction with several points $\lambda_j$ gives multi-soliton solutions
(Appendix~\ref{app:dressing}).

\subsection{Soliton}
\label{sec:onesoliton}
For one point $\lambda_1=\xi+\ii\eta$, Eq.~\eqref{eq:readoff} gives, with $R=|z_{n,m}|^2$,
\begin{equation}
\begin{aligned}
  E&=\frac{8\eta\,\bar z\,\bigl[\eta(R-1)-(1-\ii\xi)(1+R)\bigr]}{(1+R)\,\Delta_U},\qquad
  C=-\frac{8\eta^2R}{(1+R)\,\Delta_U},\\
  N&=N_0\,\frac{N_0^2(1+R)^2-8N_0\eta(R^2-1)+16\xi^2(1+R)^2+16\eta^2(R^2-6R+1)}{(1+R)\,\Delta_V},\\
  \rho&=\frac{16N_0\eta\,\bar z\,\bigl[N_0(1+R)-4\eta(R-1)-4\ii\xi(1+R)\bigr]}{(1+R)\,\Delta_V},
\end{aligned}
\label{eq:soliton}
\end{equation}
where $\Delta_U=(1+\xi^2+\eta^2)(1+R)-2\eta(R-1)$ and
$\Delta_V=N_0^2(1+R)-8N_0\eta(R-1)+16(\eta^2+\xi^2)(1+R)$.
These expressions satisfy~\eqref{eq:lattice} identically. The pulse is centered where
$|q|\,|z_{n,m}|^2=1$ and moves along the lattice line
$(n+\tfrac12)\log|q|+m\log|p|=-\log|\gamma|$. For $\xi=0$ and $\eta<1/\sqrt2$ the field obeys
$|E|\le4\eta\sqrt{1-\eta^2}$ on every time edge, with equality only if an edge lies at the
center (for suitable $\gamma$); in physical units the amplitude tends to $4\eta_{\rm ph}$. For
$\xi=0$ and $|N_0|<4\eta$ the inverse velocity in physical units is
\begin{equation}
  \frac1{v}=\frac{\Delta t}{\Delta x}\,
  \frac{\operatorname{artanh}\bigl(-N_0/4\eta\bigr)}{\operatorname{artanh}\eta},
  \label{eq:velocity}
\end{equation}
with $\eta$ and $N_0$ in lattice units. With $\eta=\Delta t\,\eta_{\rm ph}$ and
$N_0=-\Delta t\Delta x$ (ground state), Eq.~\eqref{eq:velocity} tends to the McCall--Hahn
value $1/v=1/(4\eta_{\rm ph}^2)$.

The soliton is a Bargmann (reflectionless) potential of the lattice spectral problem. As
$n\to+\infty$, $z\to0$ and $\Pi\to\mathrm{diag}(1,0)$; as $n\to-\infty$, $z\to\infty$ and
$\Pi\to\mathrm{diag}(0,1)$. The Jost solution normalized at $n\to-\infty$,
$\mathcal D\Psi^{(0)}\mathrm{diag}(1,\tau^{-1})$ with
$\tau=(\lambda-\lambda_1)/(\lambda-\bar\lambda_1)$, therefore tends to
$\Psi^{(0)}\mathrm{diag}(\tau,\tau^{-1})$ at $n\to+\infty$, and the scattering coefficients
(the entries of the scattering matrix in the first column) are
\begin{equation}
  a(\lambda)=\frac{\lambda-\lambda_1}{\lambda-\bar\lambda_1},\qquad b(\lambda)\equiv0
  \quad(\lambda\in\mathbb R),
  \label{eq:bargmann}
\end{equation}
the same as for the Zakharov--Shabat soliton of the continuous problem.

From~\eqref{eq:soliton},
$C_{n,m}=-2\eta\bigl[(1+|q|^2R_{n,m})^{-1}-(1+R_{n,m})^{-1}\bigr]$, so the sum
in~\eqref{eq:energy} telescopes and the lattice energy is $W=16\eta$ (in physical units
$16\eta_{\rm ph}$, the continuum value of $\int|E|^2dt$). For a real soliton ($\xi=0$, real
$\gamma$, $\eta<1/\sqrt2$) the discrete area $2\sum_n\alpha_n$ with $\sin\alpha_n=E_n/2$ equals
$-2\pi\,\mathrm{sgn}\,\gamma$ exactly. In the continuum limit (for $\xi=0$),
$z\to\gamma\exp(-2\eta_{\rm ph}t+x/2\eta_{\rm ph})$ and, up to a constant phase factor,
$E\to4\eta_{\rm ph}\,\mathrm{sech}(2\eta_{\rm ph}t-x/2\eta_{\rm ph}+\mathrm{const})$, the
self-induced-transparency soliton; the lattice solution converges to it with second-order
accuracy.

\subsection{Breather}
\label{sec:breather}
Two-fold dressing (Appendix~\ref{app:dressing}) with $\lambda_{1,2}=\pm\xi+\ii\eta$ and
$\gamma_2=-\bar\gamma_1$ gives a bound pair. Write $z_1=e^{-X+\ii\Phi}$ with
\begin{equation}
  X_{n,m}=X_0+n\nu_t+m\nu_x,\quad \Phi_{n,m}=\Phi_0+n\omega_t+m\omega_x,\qquad
  q=e^{-\nu_t+\ii\omega_t},\quad p=e^{-\nu_x+\ii\omega_x},
  \label{eq:XPhi}
\end{equation}
where $q,p$ are given by~\eqref{eq:zqp} at $\lambda_1$. The field and the medium then take
the explicit form
\begin{equation}
\begin{aligned}
  E_{n,m}&=2\ii\sin\frac{\theta_{n+1,m}-\theta_{n,m}}{2},&\qquad
  1+C_{n,m}&=\cos\frac{\theta_{n+1,m}-\theta_{n,m}}{2},\\
  N_{n,m}&=N_0\cos\Theta_{n,m},&\qquad \rho_{n,m}&=\ii N_0\sin\Theta_{n,m},
\end{aligned}
  \label{eq:breather}
\end{equation}
with
\[
  \theta_{n,m}=4\arctan\frac{\xi\cosh X_{n,m}}{\eta\cos\Phi_{n,m}},\qquad
  \Theta_{n,m}=\tfrac12\bigl(\theta_{n,m+1}+\theta_{n,m}\bigr).
\]
Up to the constant factor $\ii$ (a fixed carrier phase) the field and the polarization are
real. The variable $\theta$ is the angle of the discrete sine-Gordon equation of
Sec.~\ref{sec:structure}: up to the factor $\ii$, the field angle on a time edge is
$\alpha_n=(\theta_{n+1}-\theta_n)/2$, and for $N_0=-r$ the Bloch angle on a space edge is
$\Theta+\pi$. The
real case of~\eqref{eq:breather} is therefore the breather of the Hirota--Orfanidis
lattice~\cite{Hirota1977c,Orfanidis1978}; the new element is its reading as an optical
pulse in a two-level medium, together with the medium state $N,\rho$.

The two components have equal $|q|$ and $|p|$, so the envelope $\cosh X$ moves along the
lattice line $X=\mathrm{const}$ with a single velocity, while the phase $\Phi$ produces the
internal oscillation (breathing). The discrete area $2\sum_n\alpha_n$ of the pulse vanishes
(a $0\pi$ pulse). Its lattice energy is $W=16(\eta_1+\eta_2)=32\eta$; more generally,
$W=16\sum_j\eta_j$ for every multi-soliton solution, since
$C_n=-\ii[(\mathcal D_1)_{11}(n+1)-(\mathcal D_1)_{11}(n)]$, where
$\mathcal D=I+\mathcal D_1/\lambda+\dots$, and $(\mathcal D_1)_{11}$ tends to
$-2\ii\sum_j\eta_j$ and $0$ at $n\to\pm\infty$. In the continuum limit
$X=2\eta_{\rm ph}t+\dots$ and $\Phi=2\xi_{\rm ph}t+\dots$, and~\eqref{eq:breather} becomes
$E=\ii\,\partial_t\theta$ with $\theta=4\arctan[\xi\cosh X/(\eta\cos\Phi)]$, which is Lamb's
$0\pi$ breather of self-induced transparency: up to an additive constant and a shift of $\Phi$,
$\theta$ is the angle $\sigma=4\tan^{-1}[(\alpha/\beta)\sin q/\cosh p]$ of Eq.~(6.36) of
Ref.~\cite{Lamb1971} with $\alpha=\eta$, $\beta=\xi$, $p=X$ and $q=\Phi-\pi/2$ ($\theta=\sigma+2\pi$ for $|\Phi|<\pi/2$).
The lattice breather converges
to it with second-order accuracy.

Multi-soliton solutions follow by iterating the dressing. For two arbitrary points (complex
fields), $E_{n,m}=2\ii(\Xi_{n+1,m}+\Xi_{n,m})$ with $\Xi=\sum_{jk}(M^{-1})_{kj}\bar z_j$ and
$M_{jk}=(1+\bar z_jz_k)/(\lambda_k-\bar\lambda_j)$ (Appendix~\ref{app:dressing}); for the
breather $\Xi$ is real, and in the continuum limit $4\Xi/\Delta t\to\partial_t\theta$.

\section{Numerical performance}
\label{sec:numerics}
We now use the lattice as a numerical scheme: we compare its accuracy and cost with standard schemes, describe the tails left by nonsoliton pulses, and calibrate the input for solitons.

\subsection{Schemes and test problems}
\label{sec:schemes}
We compare the integrable lattice with two standard schemes that also conserve the length of
the Bloch vector exactly. All three use the same staggered grid ($h=\Delta t=\Delta x$) and the
same Maxwell step~\eqref{eq:lattice1}; they differ only in the Bloch step across a time edge,
which is driven by the field $\bar E=\tfrac12(E_{n,m}+E_{n,m+1})$:
(I) the integrable step~\eqref{eq:qstep}, with the input mapped as $E_{\rm lat}=hE(t_{n+1/2})$;
(C) the Cayley step $\hat\rho\to Q\hat\rho Q^{-1}$, $Q=(I-\tfrac h2A)^{-1}(I+\tfrac h2A)$,
$A=-\tfrac12\bar u$, where $\bar u$ is $u$ with $E$ replaced by $\bar E$, i.e., the implicit midpoint (Cayley) rule~\cite{HairerLubichWanner2006}
applied to the spinor whose projector gives $\hat\rho$; and
(S) the exponential midpoint step $Q=e^{hA}$, an exact rotation of the Bloch
vector~\cite{AllenEberly1975}, which is the building block of the splitting scheme
of Ref.~\cite{BidegarayBourgeadeReignier2001}. Schemes C and S are implicit in
$E_{n,m+1}$ and are solved by fixed-point iteration on $E_{n,m+1}$, starting from the predictor
below and stopped when the change of $E_{n,m+1}$ falls below $10^{-15}$ in absolute value (at
most 60 sweeps); scheme I is explicit. All schemes were coded in the same compiled language and
timed on one core of a desktop processor. We also
use the cheapest variants of C and S, denoted C$_1$ and S$_1$, which perform a single sweep of this
iteration: the predictor $E_{n,m+1}=E_{n,m}+\rho_{n,m}$, one Bloch step driven by
$\bar E$, and the corrector~\eqref{eq:lattice1}. They also rotate the Bloch vector exactly. The
test problems are the exact $2\pi$ soliton $E=4\eta\,\mathrm{sech}[2\eta(t-x/4\eta^2)]$ with
$\eta=0.5$, propagated to $x=10$ and, for long-distance behavior, to $x=200$, and the
nonsoliton input $E(0,t)=1.5\,\mathrm{sech}\,t$, whose reference solution is obtained by
Richardson extrapolation of scheme I at $h=0.0125$ and $0.00625$ (at the finer step it agrees
with scheme C to $10^{-6}$). All runs use the time window $t\in[-15,25)$ ($[-15,230)$ for
$x=200$); the soliton position is the maximum of a parabola through the three largest samples of
$|E|$, and for the generic input the fields are compared on the grid $t=-14+0.05k$ by linear
interpolation.

\subsection{Accuracy and cost}
\label{sec:accuracy}
Table~\ref{tab:compare} summarizes the results. All schemes converge with second order.
For the generic input the errors of I, C and S agree to within $1.1\%$ ($2.2\times10^{-4}$ at $h=0.025$).
For the soliton, scheme I is $3.7$--$5.2$ times less accurate than C and S at equal $h$. The
amplitude is not the cause: after $x=200$ at $h=0.1$ the relative peak error is $4.3\times10^{-4}$ for I
against $1.0\times10^{-3}$ for C and $5.7\times10^{-4}$ for S. The error comes from the
velocity: the soliton's position error is $-0.67$ for I against $+0.17$ for C and $-0.11$ for S.
The ranking depends on the soliton: at $h=0.05$ and $\eta=0.25$ scheme I is more accurate than
C and S ($5.7\times10^{-3}$ against $8.4\times10^{-3}$ and $7.5\times10^{-3}$), while at $\eta=1$
it is $3$--$9$ times less accurate.
Per node, scheme I is the cheapest: it requires one $2\times2$ similarity transformation, whereas
C and S iterate. Iterated to $10^{-15}$ ($3.5$--$5$ iterations per node), they are
$5$--$8$ times slower than I with identical implementations (0.013\,s versus 0.073\,s and
0.067\,s for $6.4\times10^5$ nodes at $h=0.025$; a factor of $7$--$8$ at $h=0.1$ over $x=200$). Their
error does not change if the iteration is stopped at $10^{-8}$ or after three sweeps; the factor
is then $2.5$--$5$. A single sweep is enough for second order: C$_1$ and S$_1$ cost $1.6$--$1.7$
times as much as I per node and are more accurate, by factors of about $2$ (C$_1$) and $9$ (S$_1$)
for the soliton and $2.5$ for the generic input at equal $h$. At equal accuracy S$_1$ is therefore
the cheapest of the schemes compared: for the soliton it needs less than one fifth of the time
of I, and for the generic input both C$_1$ and S$_1$ need about two thirds. The integrable
lattice thus does not gain in cost per accuracy. Its advantages are of a different kind: the
step is explicit and needs no stopping rule, the Bloch length and the energy balance hold
exactly, and the solitons and breathers of Sec.~\ref{sec:soliton} are exact solutions of the
scheme. The velocity error of I is removed in Sec.~\ref{sec:calibration}.

\begin{table}[t]
\caption{Maximal error at $x=10$ (of $|E|$ for the soliton, of the complex field $E$ for the
generic input) for the integrable (I), Cayley (C) and exponential midpoint (S) schemes iterated to
$10^{-15}$, and for their single-sweep variants C$_1$ and S$_1$, $h=\Delta t=\Delta x$; time for
$6.4\times10^5$ nodes at $h=0.025$ (for C$_1$, S$_1$ from the measured cost ratio to I; not measured for the generic input). Lower values are better; I is the scheme of this work.}
\label{tab:compare}
\begin{ruledtabular}
\begin{tabular}{lcccccc}
 & $h=0.2$ & $0.1$ & $0.05$ & $0.025$ & order & time (s)\\
\hline
soliton, I & $1.35\times10^{-1}$ & $3.35\times10^{-2}$ & $8.34\times10^{-3}$ & $2.09\times10^{-3}$ & 2.00 & 0.013\\
soliton, C & $3.51\times10^{-2}$ & $8.97\times10^{-3}$ & $2.25\times10^{-3}$ & $5.64\times10^{-4}$ & 2.00 & 0.073\\
soliton, S & $2.58\times10^{-2}$ & $6.43\times10^{-3}$ & $1.61\times10^{-3}$ & $4.01\times10^{-4}$ & 2.00 & 0.067\\
soliton, C$_1$ & $6.87\times10^{-2}$ & $1.75\times10^{-2}$ & $4.41\times10^{-3}$ & $1.10\times10^{-3}$ & 2.00 & 0.022\\
soliton, S$_1$ & $1.48\times10^{-2}$ & $3.74\times10^{-3}$ & $9.38\times10^{-4}$ & $2.35\times10^{-4}$ & 2.00 & 0.021\\
generic, I & $1.40\times10^{-2}$ & $3.47\times10^{-3}$ & $8.66\times10^{-4}$ & $2.17\times10^{-4}$ & 2.00 & --- \\
generic, C & $1.41\times10^{-2}$ & $3.50\times10^{-3}$ & $8.72\times10^{-4}$ & $2.18\times10^{-4}$ & 2.00 & --- \\
generic, S & $1.41\times10^{-2}$ & $3.49\times10^{-3}$ & $8.70\times10^{-4}$ & $2.17\times10^{-4}$ & 2.00 & --- \\
generic, C$_1$ & $5.63\times10^{-3}$ & $1.41\times10^{-3}$ & $3.50\times10^{-4}$ & $8.75\times10^{-5}$ & 2.00 & --- \\
generic, S$_1$ & $5.59\times10^{-3}$ & $1.39\times10^{-3}$ & $3.47\times10^{-4}$ & $8.66\times10^{-5}$ & 2.00 & --- \\
\end{tabular}
\end{ruledtabular}
\end{table}

\begin{figure}[t]
\includegraphics[width=\columnwidth]{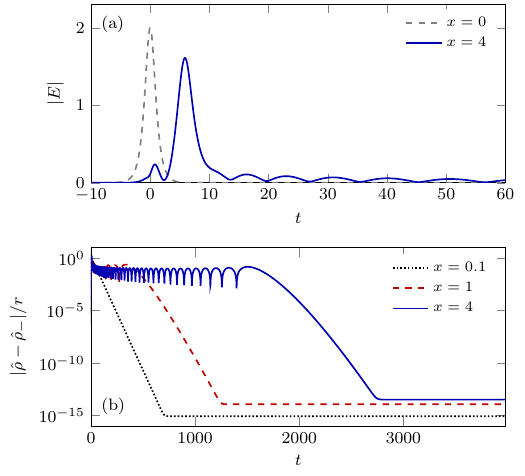}
\caption{Nonsoliton input $E(0,t)=2\,\mathrm{sech}(t/0.8)\,e^{0.3\ii t}$, $h=0.1$. (a)~$|E|$ at
$x=0$ and $x=4$: the pulse reshapes and leaves a ringing tail. (b)~Deviation of the medium from
the ground state behind the pulse at $x=0.1$, $1$ and $4$. On the first row it decays at the rate
$\Delta x/2$ per unit time (Sec.~\ref{sec:linearR}); on later rows the ringing lasts longer and,
over the times shown, the return is slower. The plateaus are round-off.}
\label{fig:tail}
\end{figure}

\label{sec:tails}
The exact solitons of Sec.~\ref{sec:soliton} propagate without shedding radiation: after 200
steps the field at the edges of the time window stays below $10^{-100}$, and $W=16\eta$ holds to
$10^{-13}$. The scheme itself reproduces the closed-form soliton~\eqref{eq:soliton} to
$9\times10^{-14}$ after 200 steps and the breather~\eqref{eq:breather} to $1.4\times10^{-14}$. A pulse that is not a
pure soliton reshapes and leaves a slowly decaying, nonmonotonic ringing polarization behind
it (Fig.~\ref{fig:tail}). For $E(0,t)=2\,\mathrm{sech}(t/0.8)\,e^{0.3\ii t}$, $|t|\le20$, at
$h=0.1$ and $x=4$, the deviation of the medium from the ground state,
$|\hat\rho-\hat\rho_-|/r\equiv[|\rho|^2+(N-N_-)^2]^{1/2}/r$, equals $0.20$, $0.11$ and $0.06$ at the end of the input
and $100$ and $400$ time units later; the medium reaches the ground state to $10^{-12}$ only at
$t\approx2.6\times10^3$, by the mechanism described in Sec.~\ref{sec:linearR}. This late return
is a lattice effect whose time scales as $1/\Delta x$: at $h=0.05$ the envelope of the deviation
agrees with that at $h=0.1$ to about $15\%$ up to $t\approx10^3$, but the medium has not yet
returned at the end of the window, $t=4\times10^3$.

The energy balance~\eqref{eq:energy} holds to round-off ($3\times10^{-17}$ in a test run).

\label{sec:dispersion}
Linearizing~\eqref{eq:lattice} about $E=\rho=0$, $N=N_0$ with $E,\rho\propto\mu^n\nu^m$ gives
\begin{equation}
  (\nu-1)(\mu-1)=\tfrac{N_0}{4}(\nu+1)(\mu+1).
  \label{eq:dispersion}
\end{equation}
For a temporal Fourier mode, $|\mu|=1$, this yields $|\nu|=1$ for both signs of $N_0$: small
signals propagate neither amplified nor damped, in the absorbing as well as in the inverted
medium, as in the continuous signaling problem, where $E_{xt}=N_0E$ gives a purely oscillatory
$x$-dependence for real frequencies. In the Cayley variables
$\nu=(1+\ii\Delta x\,k/2)/(1-\ii\Delta x\,k/2)$ and
$\mu=(1-\ii\Delta t\,\omega/2)/(1+\ii\Delta t\,\omega/2)$, Eq.~\eqref{eq:dispersion} becomes exactly
the continuous relation $k\omega=N_0$.

\subsection{Calibrated input}
\label{sec:calibration}
\begin{figure}[t]
\includegraphics[width=\columnwidth]{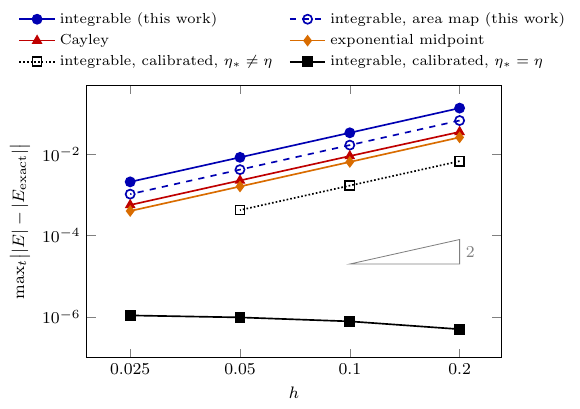}
\caption{Maximal error of $|E|$ at $x=10$ for the $2\pi$ soliton with $\eta=0.5$ versus the step
$h=\Delta t=\Delta x$: integrable scheme with point sampling and with the input
map~\eqref{eq:areamap}, Cayley and exponential midpoint schemes, and the integrable scheme with the
map~\eqref{eq:areamap} and the calibration~\eqref{eq:calibration} for $\eta_*=\eta$ and for a
soliton with $\eta=0.4$ at $\eta_*=0.5$. With the input map the output is read back by its
inverse and compared with the cell averages of the exact field. The triangle marks the slope $2$; the floor near $10^{-6}$ of the calibrated scheme is set by the time window (Sec.~\ref{sec:calibration}).}
\label{fig:accuracy}
\end{figure}

The velocity error of scheme I has two sources. First, sampling $E_{\rm lat}=hE(t_{n+1/2})$
does not place an incoming soliton exactly on a lattice soliton. The natural input map
preserves the pulse area,
\begin{equation}
  E_{{\rm lat},n}=2\sin\frac{|A_n|}{2}\,e^{\ii\arg A_n},\qquad A_n=\int_{t_n}^{t_{n+1}}E\,dt,
  \label{eq:areamap}
\end{equation}
with the output field read back as $2\arcsin(|E_{\rm lat}|/2)\,e^{\ii\arg E_{\rm lat}}/h$. For
real fields~\eqref{eq:areamap} samples exactly the sine-Gordon angle $\theta(t_n)$ of
Sec.~\ref{sec:breather}, so that the discrete area equals the continuous one, and
$|E_{\rm lat}|\le2$, so that the normalization~\eqref{eq:normalization} is always defined. For the
soliton, \eqref{eq:areamap} gives exactly the lattice soliton with $\eta_{\rm lat}=\tanh(\eta h)$,
because $|q|=e^{-2\eta h}$. Second, by~\eqref{eq:velocity} the velocity of this soliton has the
relative error $\tfrac13[(\eta h)^2+(\Delta x/4\eta)^2]+O(h^4)$. The map~\eqref{eq:areamap} alone
halves the error (Fig.~\ref{fig:accuracy}); the remaining position error after $x=200$ at
$h=0.1$, 0.33403,
agrees with the value 0.33397 predicted by~\eqref{eq:velocity}. The medium enters only through
$r$ in $p$, so the choice
\begin{equation}
  r=4\tanh(\eta_*h)\tanh(\Delta x/4\eta_*),
  \label{eq:calibration}
\end{equation}
which equals $h\Delta x+O(h^3)$, makes the $x$-factor $|p|$ exact for $\eta=\eta_*$. The lattice
then reproduces the continuous soliton with $\eta_*$ exactly at any $h$: the remaining error,
$10^{-6}$, comes from the truncation of the time window (widening the window from $[-15,25)$ to
$[-30,40)$ lowers it below $10^{-12}$), and the position error after $x=200$ is
$2\times10^{-8}$. A single medium constant cannot be exact for all $\eta$, so other solitons keep
an $O(h^2)$ error, but a small one: for $\eta=0.4$ and $0.6$ with $\eta_*=0.5$ the errors at
$h=0.05$ are $4.2\times10^{-4}$ and $2.8\times10^{-4}$, $2.6$ to $13$ times smaller than those of
C ($3.3\times10^{-3}$ and $2.0\times10^{-3}$) and S ($1.1\times10^{-3}$ and $3.7\times10^{-3}$),
iterated to convergence, for the same solitons. For generic pulses, the map~\eqref{eq:areamap}
alone changes the error by only a few percent ($2.11\times10^{-4}$ versus $2.17\times10^{-4}$ at
$h=0.025$), but the calibration tuned to a soliton increases it by a factor of $1.7$
($3.58\times10^{-4}$); the calibration is therefore useful only when the pulses of interest are
close to the chosen soliton. It has been tested for real solitons ($\xi=0$).

\section{Inhomogeneous broadening}
\label{sec:broadening}
For an inhomogeneously broadened line the direct lattice analogue of the continuous Lax pair fails; an ordered product of single-class steps keeps integrability.

\subsection{Model and the failure of the additive lattice}
\label{sec:broad-model}
In an inhomogeneously broadened medium the atoms have a distribution $g(\epsilon)$ of detunings,
$\int g(\epsilon)\,d\epsilon=1$. Each detuning class has its own density matrix $\hat\rho(x,t,\epsilon)$, and
the field is driven by their average~\cite{AllenEberly1975,GabitovZakharovMikhailov1985}:
\begin{equation}
  E_x=\int g(\epsilon)\,\rho(\epsilon)\,d\epsilon,\qquad \rho_t+2\ii\epsilon\rho=NE,\qquad
  N_t=-\tfrac12\bigl(E^*\rho+E\rho^*\bigr),
  \label{eq:MBbroad}
\end{equation}
with $N^2+|\rho|^2=1$ for every $\epsilon$. The sharp line of Sec.~\ref{sec:prelim} corresponds to
$g=\delta(\epsilon-\lambda_0)$; a typical example is the Lorentzian line
$g=(\gamma/\pi)/[(\epsilon-\epsilon_0)^2+\gamma^2]$. Equations~\eqref{eq:MBbroad} keep the Lax
structure~\eqref{eq:lax} with the same $U$ and
\begin{equation}
  V=-\frac{1}{4\ii}\int\frac{g(\epsilon)\,\hat\rho(\epsilon)}{\lambda-\epsilon}\,d\epsilon,
  \label{eq:Vbroad}
\end{equation}
so that each detuning class contributes a simple pole at $\lambda=\epsilon$~\cite{GabitovZakharovMikhailov1985}.
For a continuous $g$ these poles merge into a cut along the real axis, which in the continuous
inverse scattering theory requires a limiting prescription
($\epsilon-\lambda\pm\ii0$)~\cite{GabitovZakharovMikhailov1985,AbeyaBiondiniKovacicPrinari2024}.
On the lattice the cut is replaced by a product of single-pole factors, one per detuning class.

\label{sec:additive}
The direct lattice analogue of~\eqref{eq:Vbroad} places all classes on the same space edge,
$\mathbf V_{n,m}=I-\sum_jw_j\hat\rho_{j;n,m}/[4\ii(\lambda-\epsilon_j)]$, where $\epsilon_j$ and $w_j$ are
the nodes and weights of a quadrature of $g$. This does not give an integrable lattice. The
residue of~\eqref{eq:compat} at $\lambda=\epsilon_j$ fixes
$\hat\rho''_j=\mathbf U_{n,m+1}(\epsilon_j)\,\hat\rho_j\,\mathbf U_{n,m}(\epsilon_j)^{-1}$. Since
$\mathbf U(\epsilon)/\sqrt{1+\epsilon^2}$ is unitary for real $\epsilon$, $\hat\rho''_j$ is Hermitian only if
$\mathbf U_{n,m+1}(\epsilon_j)\hat\rho_j\mathbf U^\dagger_{n,m}(\epsilon_j)
=\mathbf U_{n,m}(\epsilon_j)\hat\rho_j\mathbf U^\dagger_{n,m+1}(\epsilon_j)$. For a single class this
condition determines $E_{n,m+1}$ and leads back to~\eqref{eq:lattice}; for two or more classes
sharing one field it cannot be satisfied in general. With two classes, Eq.~\eqref{eq:compat}
still has a solution, but $\hat\rho''_j$ acquires a non-Hermitian part and a trace of the order
of the coupling $r_j\propto\Delta x$ ($1.6\times10^{-4}$ to $3.6\times10^{-3}$ for $r_j=0.006$ and
$0.004$, two classes at $\epsilon_j=0.13$ and $-0.27$, random data): the
additive lattice is consistent only to leading order in $\Delta x$.

\subsection{Ordered sublayers}
\label{sec:sublayers}
We therefore let the atoms of a cell act on the field one detuning class at a time
[Fig.~\ref{fig:lattice}(b)]. A space step
is split into $J$ sublayers $s=(m,j)$ carrying the classes $\epsilon_j$ with Bloch radii
$r_j=w_j\Delta t\Delta x$ (lattice units, $\epsilon_j\to\Delta t\,\epsilon_j$), and the $x$-step of a cell
is the ordered product
\begin{equation}
  \mathbf V_m(\lambda)=\mathbf V_{m,J}(\lambda)\cdots\mathbf V_{m,1}(\lambda),\qquad
  \mathbf V_{m,j}(\lambda)=I-\frac{\hat\rho_{m,j}}{4\ii(\lambda-\epsilon_j)},
  \label{eq:Vproduct}
\end{equation}
so that the field passes the class $j=1$ first. This product agrees with the lattice analogue
of~\eqref{eq:Vbroad} to $O(\Delta x^2)$. For a Lorentzian line we use the nodes
$\epsilon_j=\gamma\tan\vartheta_j$ on a uniform midpoint grid in $\vartheta\in(-\pi/2,\pi/2)$ with
equal weights $w_j=1/J$. Each factor is a
single-pole step, for which~\eqref{eq:compat} has the unique explicit solution (one prime marks the updated time edge, two primes the updated space edge)
\begin{equation}
  E'=E+\tfrac12(\rho+\rho''),\quad C'=C-\tfrac14(N-N''),\quad
  \hat\rho''=G_\epsilon\hat\rho\,G_\epsilon^{-1},\quad
  G_\epsilon=(1+C)I-\ii\epsilon\sigma_3-\tfrac12u-\tfrac14\sigma_3\hat\rho .
  \label{eq:detunedstep}
\end{equation}
Since $(\lambda-\epsilon_j)\mathbf V_j=(\lambda-\epsilon_j)I+\tfrac{\ii}{4}\hat\rho_j$ is a Heisenberg Lax
operator with a shifted spectral parameter, the broadened lattice is the nonautonomous form of
Tsuchida's system~\cite{Tsuchida2015}, whose parameters may depend on the lattice coordinate; here
$a_{{\rm T},s}=-\epsilon_j+\ii r_j/4$. Consequently every atom keeps the length of its Bloch vector, $\kappa$ is
conserved, and summing the $C$-equation over the sublayers gives the energy balance
$W'-W=-2\sum_j(N_{j,\rm end}-N_{j,\rm start})$, the broadened counterpart of~\eqref{eq:energy}.

\label{sec:broad-soliton}
The dressing of Sec.~\ref{sec:dressing} carries over unchanged. On the broadened background
($E=C=0$, every class in its ground state $N_j=-r_j$) the free wave function is
\[
  \Psi^{(0)}_{n,s}=\mathrm{diag}\Bigl((1-\ii\lambda)^n\prod_{s'<s}c_{s'},\;
  (1+\ii\lambda)^n\prod_{s'<s}d_{s'}\Bigr),\qquad
  c_s,d_s=1\mp\frac{N_s}{4\ii(\lambda-\epsilon_s)},
\]
and one dressing factor~\eqref{eq:dressing} gives a soliton of the broadened lattice. This
solution satisfies the sublayer step~\eqref{eq:detunedstep} at every node to $3\times10^{-15}$,
with the normalization~\eqref{eq:normalization} and every Bloch length preserved. Marching the
scheme from the soliton at $x=0$ (Lorentzian line with $\epsilon_0=0$, $\gamma=0.5$, $\lambda_1=0.4\,\ii$, $h=0.1$) through 30 cells of 8 sublayers reproduces the exact solution
to $2\times10^{-14}$, the energy balance holds on every sublayer to $4\times10^{-15}$, and the
soliton remains reflectionless, $a=(\lambda-\lambda_1)/(\lambda-\bar\lambda_1)$ and $b=0$ on the
real axis, to the accuracy set by the time window.

\label{sec:order}
The ordering in~\eqref{eq:Vproduct} is a convention, and it does not affect the field when the
medium starts in an incoherent state. We passed the chirped pulse $E(0,t)=2.2\,\mathrm{sech}(t/0.8)\,e^{0.4\ii t}$ through a
cell of six Lorentzian sublayers in the natural, reversed and random order, with the coupling
chosen so that one cell changes the field by $0.1$: the output fields coincide to
$4\times10^{-15}$, also
after ten cells. If the atoms are prepared with coherence ($\rho\neq0$), reversing the order
changes the output by $3\times10^{-2}$. The final states of individual atoms do depend on the
order, by $(1$--$2)\times10^{-2}$ in this test, because within a cell each atom sees a slightly
different field; only
the field leaving the cell is order-independent. This is the Bianchi permutability
(Yang--Baxter property) of the single-pole
steps~\cite{GoncharenkoVeselov2004,KouloukasPapageorgiou2009}, and it can be proved directly.
Let $X_n=\mathbf V_{n,J}\cdots\mathbf V_{n,1}$ be the matrix of the cell on the time edge $n$ for
one order and $\tilde X_n$ for another, with the same input $\mathbf U_n$ and outputs
$\mathbf U'_n$, $\tilde{\mathbf U}'_n$. The compatibility of the sublayer steps gives
$\mathbf U'_nX_n=X_{n+1}\mathbf U_n$ and $\tilde{\mathbf U}'_n\tilde X_n=\tilde X_{n+1}\mathbf U_n$.
If $X_n=\tilde X_n$, then $\tilde X_{n+1}X_{n+1}^{-1}=\tilde{\mathbf U}'_n\mathbf U_n'^{-1}$. The
left-hand side has poles only at $\epsilon_j$ and $\epsilon_j\pm\ii r_j/4$, the right-hand side
only at the zeros $\lambda=\pm\ii$ of $\det\mathbf U'_n$; for real $\epsilon_j$ and $r_j<4$ these
sets are disjoint, so both sides are polynomials in $\lambda$, and since they tend to $I$ as
$\lambda\to\infty$, both equal $I$. Hence $\tilde{\mathbf U}'_n=\mathbf U'_n$ and
$\tilde X_{n+1}=X_{n+1}$, and by induction the output field is the same on every time edge
whenever the ordered products agree initially. For diagonal (incoherent) initial $\hat\rho_j$ the
factors commute, so this holds for every order, in the ground state, inverted or mixed; for four
sublayers, all $24$ orders and three cells the output fields agree to $3\times10^{-16}$ in each
of these cases. The atomic states themselves are refactorized and may differ. The lattice thus
realizes a continuous line in every cell without an ordering ambiguity in the field.

\subsection{Lorentzian line}
\label{sec:lorentz}
For a Lorentzian line with $\epsilon_0=0$ the soliton with $\lambda_1=\ii\eta$ moves with the
McCall--Hahn velocity
$1/v=\int g(\epsilon)\,d\epsilon/[4(\eta^2+\epsilon^2)]$~\cite{McCallHahn1969}, which for the
Lorentzian line equals $1/[4\eta(\eta+\gamma)]$.
On the lattice the velocity follows from $|q|$ and $\prod_j|p_j|$. With the nodes of
Sec.~\ref{sec:sublayers}, $\lambda_1=0.4\,\ii$ and $\gamma=0.5$, it converges to the continuum
value with second order: the relative error is $5.3\times10^{-4}$, $1.3\times10^{-4}$,
$3.3\times10^{-5}$ and $8.3\times10^{-6}$ for $h=0.1$, $0.05$, $0.025$ and $0.0125$ (with
$J=8$, $8$, $256$ and $1024$), and $J=64$ gives the same error as $J=8$ at the two largest steps.

One might hope to avoid sublayers for this line shape: in the linear regime a Lorentzian ensemble
acts as a single atom with complex detuning, $\langle\rho\rangle=\rho(\epsilon_0-\ii\gamma)$, which
would give a single complex pole per node. In the nonlinear regime this closure fails already
for the continuous Bloch equations. For a Lorentzian line with $\epsilon_0=0.2$, $\gamma=0.5$ and a prescribed field $E=A\,\mathrm{sech}\,t$, $|t|\le12$, the relative
discrepancy is $2\times10^{-3}$ at $A=0.02$, $5.4$ at area $\pi$ and $10$ at area $2\pi$, because
the continuation of $\rho^*$ into the lower half-plane grows, so that $\rho(\epsilon)$ is not bounded
there. The ordered sublayers are therefore needed even for the Lorentzian line.

\section{Evolution of the scattering data}
\label{sec:scattering}
We express the state of the medium through the scattering data, derive their closed evolution on a finite time window, and find the linear law on the infinite line.

\subsection{Monodromy matrix and the state of the medium}
\label{sec:monodromy}
We consider a row $m$ on a time window $0\le n<n_{\max}$, with the medium before the pulse in the
ground state, $\hat\rho_{0,m}=\mathrm{diag}(-r,r)$, and define the monodromy matrix and the
normalized scattering matrix
\begin{equation}
  T(\lambda,m)=\mathbf U_{n_{\max}-1,m}(\lambda)\cdots\mathbf U_{0,m}(\lambda),\qquad
  S(\lambda,m)=\mathbf U_\infty(\lambda)^{-n_{\max}}\,T(\lambda,m)
  =\begin{pmatrix}a&\cdot\\ b&\cdot\end{pmatrix},
  \label{eq:Sdef}
\end{equation}
where $\mathbf U_\infty=I-\ii\lambda\sigma_3$ is the free $x$-independent factor of~\eqref{eq:Psi0}. For a field
supported in the window, the solution equal to $\mathbf U_\infty^n$ at the left end equals
$\mathbf U_\infty^nS$ at the right end; $S$ is thus normalized by the free evolution in $n$ only,
not by the $x$-dependent factor of~\eqref{eq:Psi0}. Its entries $a$ and $b$ are the scattering
coefficients of Sec.~\ref{sec:onesoliton}; for a soliton they approach~\eqref{eq:bargmann} as the
window grows. We call $a$ the transmission coefficient and $R=b/a$ the reflection coefficient.
Numerically, $S$ is accumulated as the product of the normalized factors
$\mathbf U_\infty^{-1}\mathbf U_{n,m}$, which avoids overflow on long windows. The compatibility condition~\eqref{eq:compat} telescopes to
\begin{equation}
  T(\lambda,m+1)=\mathbf V_{n_{\max},m}(\lambda)\,T(\lambda,m)\,\mathbf V_{0,m}(\lambda)^{-1},
  \label{eq:Tevolution}
\end{equation}
so that the coefficients of $\operatorname{tr}T$ are conserved whenever the medium after the window
has returned to its initial state.

The state of the medium is itself a function of the scattering data. At $\lambda_*=\ii r/4$, a zero
of $\det\mathbf V$, the matrix $\mathbf V_{n,m}(\lambda_*)=I+\hat\rho_{n,m}/r$ has rank one, and its
kernel is the eigenvector of $\hat\rho_{n,m}$ with eigenvalue $-r$. If
$\mathbf V_{n,m}(\lambda_*)v=0$, then~\eqref{eq:compat} gives
$\mathbf V_{n+1,m}(\lambda_*)\,\mathbf U_{n,m}(\lambda_*)v=\mathbf U_{n,m+1}(\lambda_*)\mathbf V_{n,m}(\lambda_*)v=0$,
and $\mathbf U_{n,m}(\lambda_*)v\neq0$ because $\det\mathbf U(\lambda_*)=1-r^2/16\neq0$ for $r<4$. The
kernel is therefore transported along $n$ by $\mathbf U(\lambda_*)$. Starting from $e_1=(1,0)^T$ at
$n=0$, the eigenvector at $n$ is $w=T_n(\lambda_*,m)e_1$, where $T_n$ is the product of the first $n$
factors, and since $\hat\rho$ is traceless and Hermitian with eigenvalues $\pm r$,
\begin{equation}
  \hat\rho_{n,m}=r\Bigl(I-\frac{2ww^\dagger}{w^\dagger w}\Bigr),\qquad
  w=T_n(\lambda_*,m)\,e_1 .
  \label{eq:windowGZM}
\end{equation}
In the continuum limit, $\lambda_*\to0$ and $T_n(0)$ becomes unitary, so that~\eqref{eq:windowGZM}
turns into $\hat\rho=T\hat\rho_0T^{-1}$, the sharp-line form of the relation
$\hat\rho=\Psi_0\hat\rho_0\Psi_0^{-1}$, Eq.~(1.37) of Ref.~\cite{GabitovZakharovMikhailov1985}. At the end of the pulse it becomes the
relation $\hat\rho^+=S\hat\rho^-S^{-1}$ of that work and reproduces its expressions for the final state of the
medium. The lattice evaluates this relation at the point
$\lambda_*$ in the upper half-plane, whereas in the continuous problem it is taken at the line center
on the real axis.

\subsection{Closed evolution on a finite window}
\label{sec:closedmap}
Taking $n=n_{\max}$ in~\eqref{eq:windowGZM} expresses the medium state $\hat\rho_+$ behind the window
through $S(\lambda_*,m)$, and~\eqref{eq:Tevolution} becomes a closed map
\begin{equation}
  S(\lambda,m+1)=\mathbf U_\infty^{-n_{\max}}\mathbf V_+\,\mathbf U_\infty^{n_{\max}}\,S(\lambda,m)\,\mathbf V_-^{-1},
  \qquad \mathbf V_\pm=I-\frac{\hat\rho_\pm}{4\ii\lambda},
  \label{eq:Smap}
\end{equation}
with $\hat\rho_-=\mathrm{diag}(-r,r)$ and $\hat\rho_+$ given by~\eqref{eq:windowGZM} with
$w=\mathbf U_\infty(\lambda_*)^{n_{\max}}S(\lambda_*,m)e_1$. The step needs $S$ at $\lambda$ and at the
single point $\lambda_*$. For the reflection coefficient it is a M\"obius map,
\begin{equation}
  R'=\frac{\bigl(1+\frac{N_+}{4\ii\lambda}\bigr)R-\frac{\bar\rho_+}{4\ii\lambda}Q^{-n_{\max}}}
          {1-\frac{N_+}{4\ii\lambda}-\frac{\rho_+}{4\ii\lambda}Q^{n_{\max}}R},
  \qquad Q(\lambda)=\frac{1+\ii\lambda}{1-\ii\lambda},
  \label{eq:Rmap}
\end{equation}
where $R'=R(\lambda,m+1)$, $R=R(\lambda,m)$, and the final state of the medium is
\begin{equation}
  N_+=r\,\frac{|\zeta|^2-1}{|\zeta|^2+1},\qquad
  \rho_+=-\frac{2r\bar\zeta}{1+|\zeta|^2},\qquad
  \zeta=Q(\lambda_*)^{n_{\max}}R(\lambda_*,m),\quad Q(\lambda_*)=\frac{1-r/4}{1+r/4}.
  \label{eq:finalstate}
\end{equation}
To first order in $r$, Eq.~\eqref{eq:Rmap} is the discrete Riccati equation
$R'-R=\bigl(2N_+R-\bar\rho_+Q^{-n_{\max}}+\rho_+Q^{n_{\max}}R^2\bigr)/(4\ii\lambda)$. Together
with~\eqref{eq:finalstate} it is the lattice counterpart of Eqs.~(1.28)--(1.29) of
Ref.~\cite{GabitovZakharovMikhailov1985} closed by its expressions~(1.39)--(1.40) for the final
state of the medium.

For inhomogeneous broadening the map is applied sublayer by sublayer in the order
of~\eqref{eq:Vproduct}: in the step through the class $\epsilon_j$ one replaces $\lambda$ by
$\lambda-\epsilon_j$ in $\mathbf V_\pm$ and in the denominators $4\ii\lambda$ of~\eqref{eq:Rmap}, $r$ by
$r_j$, and $\lambda_*$ by $\lambda_{*,j}=\epsilon_j+\ii r_j/4$; the factors $Q^{\pm n_{\max}}$ keep the
argument $\lambda$, while $\zeta$ in~\eqref{eq:finalstate} uses the complex factor
$Q(\lambda_{*,j})$ (the real value given there is that of the sharp line). We checked Eqs.~\eqref{eq:windowGZM}--\eqref{eq:finalstate} against the explicit
step in 60-digit arithmetic, for a sharp line and for three sublayers, at real and complex
$\lambda$: the residuals are below $10^{-45}$.

\subsection{Infinite line: linear evolution}
\label{sec:linearR}
Let the field decay behind the pulse. Then $\mathbf U_{n,m}(\lambda_*)\to\mathbf U_\infty(\lambda_*)=
\mathrm{diag}(1+r/4,\,1-r/4)$, and the second component of $w$ in~\eqref{eq:windowGZM} decays
relative to the first by the factor $Q(\lambda_*)$ per step, provided $a(\lambda_*)\neq0$. The medium
therefore returns to the ground state behind the pulse, $\zeta\to0$ in~\eqref{eq:finalstate}, at the
rate $\ln[(1+r/4)/(1-r/4)]\simeq r/2$ per time step, that is $\Delta x/2$ per unit time in physical
units. On the first row the measured rate agrees with $r/2$ to four digits [Fig.~\ref{fig:tail}(b)]. On later rows the field
behind the pulse carries a tail inherited from the polarization of the previous rows, and the window
needed for the return grows with $m$; we attribute this to a decay of the form
$P_m(n)\,Q(\lambda_*)^n$ with a polynomial $P_m$, but have not derived it. This
return of the medium on the lattice is the counterpart of the ringing polarization of
Sec.~\ref{sec:accuracy}, which persists over finite windows.

As $n_{\max}\to\infty$, for real $\lambda$ and in the strip
$|Q(\lambda_*)|<|Q(\lambda)|<|Q(\lambda_*)|^{-1}$ around the real axis (of width about $r/4$), the
map~\eqref{eq:Smap} reduces to $S(m+1)=\mathbf V_-S(m)\mathbf V_-^{-1}$,
and the scattering data evolve linearly:
\begin{equation}
  a(\lambda,m+1)=a(\lambda,m),\qquad R(\lambda,m+1)=\chi(\lambda)\,R(\lambda,m),\qquad
  \chi(\lambda)=\prod_{j}\frac{\lambda-\epsilon_j+\ii r_j/4}{\lambda-\epsilon_j-\ii r_j/4},
  \label{eq:linearR}
\end{equation}
where the product has the single factor $(\lambda+\ii r/4)/(\lambda-\ii r/4)$ for a sharp line. The
same factor $\chi(\lambda_1)=p$ appears in~\eqref{eq:zqp} (where also $q=Q(\lambda_1)$), so for a pure soliton the norming
constant evolves by the same law. The soliton eigenvalues lie outside the strip, but the law holds
for them as well in our test: for the input $1.5\,\mathrm{sech}\,t$ below, which contains one
soliton, the eigenvalue $\lambda_1=0.0250\,\ii$ stays fixed and the norming constant changes over
40 steps by $p^{40}$ to a relative accuracy of $10^{-12}$. The multiplier $\chi$ is the multiplier of linear waves on the lattice:
with $\mu=Q(\lambda)$, i.e. $\lambda=-\Delta t\,\omega/2$, the dispersion
relation~\eqref{eq:dispersion} gives exactly $\nu=\chi(\lambda)$. Hence the superposition
$F_{n,m}=\int R(\lambda,0)\,\chi(\lambda)^m Q(\lambda)^n\,d\lambda$ satisfies the linearized lattice
equations. This is the discrete analogue of Proposition~2 of Ref.~\cite{GabitovZakharovMikhailov1985}, by which the Fourier transform of $c=b^*/a$ obeys
the linearized Maxwell--Bloch equations.

\begin{figure}[t]
\includegraphics[width=\columnwidth]{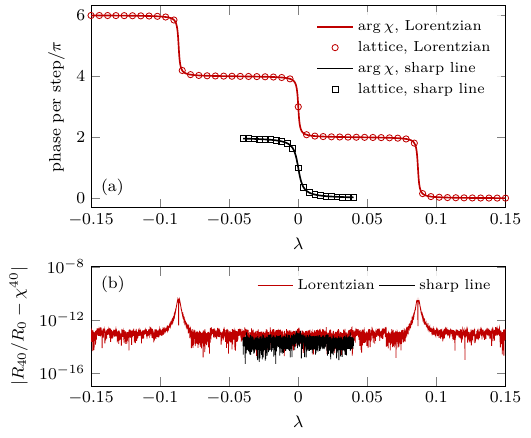}
\caption{Linear evolution of the reflection coefficient on the infinite line, input
$1.5\,\mathrm{sech}\,t$, $h=0.1$, after $m=40$ steps in $x$; sharp line and Lorentzian line
($\gamma=0.5$ in physical units) with three sublayers at $\epsilon_j=0,\pm0.087$ (lattice units, like
$\lambda$; $\pm0.87$ in physical units). (a)~Phase per step of
$R(\lambda,40)/R(\lambda,0)$ computed on the lattice and continued in $\lambda$ (symbols), and
$\arg\chi(\lambda)$ from~\eqref{eq:linearR} (lines). Each resonance adds $2\pi$. (b)~Deviation
from~\eqref{eq:linearR}; the peaks at $\epsilon_j=\pm0.087$, where the phase of $\chi$ varies
fastest, do not change when the time window is doubled.}
\label{fig:linearR}
\end{figure}

In the checks below the input starts at $t=-15$; the soliton eigenvalue $\lambda_1$ is found by
bisection as the zero of $a$ on the imaginary axis, and its norming constant $\beta$ from the
proportionality $\phi=\beta\psi$ of the Jost solutions normalized at the two ends of the window.
We checked~\eqref{eq:linearR} for the input $1.5\,\mathrm{sech}\,t$ at $h=0.1$, which contains
one soliton ($\eta_{\rm ph}=0.250$) and radiation, at nine points $|\lambda|\le0.4$ (lattice units;
points with $|R|<10^{-6}$ are excluded) and in the resonant region $|\lambda|\le r$, where
$|\chi^m-1|$ reaches $2$. After $m=10$ and $m=40$ steps in
$x$, $R(\lambda,m)/R(\lambda,0)$ equals $\chi^m$ to $3\times10^{-11}$, and the deviation of the
medium behind the window from the ground state, $|\hat\rho_+-\hat\rho_-|/r$, is below
$10^{-13}$, once the window exceeds $1.6\times10^4$ and $3.2\times10^4$ time steps, respectively;
on windows of $4\times10^3$ and $1.6\times10^4$ steps it is still $0.25$ and $0.16$. For a
Lorentzian line with three sublayers, on a window of $1.3\times10^5$ steps, the error is
$5\times10^{-11}$ and the deviation $4\times10^{-12}$; the longer window is needed because each
class relaxes at its own, smaller rate, $r_j/[2(1+\epsilon_j^2)]\approx r_j/2$. Figure~\ref{fig:linearR} shows the comparison on a dense grid in $\lambda$ after $m=40$
steps: the phase of $R(\lambda,40)/R(\lambda,0)$ per step, continued in $\lambda$, coincides with
$\arg\chi$ to $10^{-12}$, and $|R(\lambda,40)/R(\lambda,0)-\chi^{40}|\le4\times10^{-11}$.

On the lattice, then, $a$ is conserved and $R$ evolves linearly. For the real reduction, Hirota's
discrete sine-Gordon equation in Orfanidis' Lax pair, the same law was obtained by Boiti
\textit{et al.}~\cite{BoitiPempinelliPrinariSpire2002}, whose Eqs.~(6.38)--(6.39) give a conserved
$a$ and a factor for $b$ that, with $\gamma_{\rm O}=r/4$ and $k_{\rm O}=-\lambda$, equals $\chi^{\pm1}$
depending on the convention for $b$; Eq.~\eqref{eq:linearR} extends it to complex fields and
broadened lines. In the continuous sharp-line
problem only $c=b^*/a$ evolves linearly [for a medium initially in the ground state it obeys $\partial_xc=\ii\varphi(x,\lambda)c$, Eq.~(1.30) of Ref.~\cite{GabitovZakharovMikhailov1985}], while $a$ changes according to Eq.~(1.28) of
Ref.~\cite{GabitovZakharovMikhailov1985}, because the medium keeps a polarization after the pulse.
We attribute the difference to the order of limits: the relaxation time $2/\Delta x$ diverges as
$\Delta x\to0$, so the lattice result holds on the infinite line at fixed $\Delta x$. The lattice
moves the resonance from the real axis to the points $\epsilon_j\pm\ii r_j/4$; this shift takes the role of
the $\pm\ii0$ prescription in the continuous problem. How the continuum limit of~\eqref{eq:linearR}
recovers the continuous evolution, and how resonant absorption of a broadened line arises from it as
$J\to\infty$, we leave open.

\section{Conclusions}
\label{sec:discussion}
We have written an integrable full discretization of the Maxwell--Bloch equations for a
complex field in the physical variables and used it as a numerical scheme. Its key properties hold exactly on the lattice: the step is explicit, every Bloch
vector and the field spin keep their lengths, the energy balance between field and medium is
exact, and the solitons and breathers are exact reflectionless solutions. The connection with
Tsuchida's fully discrete system was our main tool; it gives the exact solutions and the
extension to inhomogeneously broadened media by ordered sublayers, whose order does not affect the
field when the medium starts in an incoherent state. As a numerical method the
lattice is explicit and second order; its cost per accuracy is comparable to that of Cayley and
exponential midpoint schemes, a single sweep of which is already second order, and with a
calibrated input it is exact for a chosen soliton. For the scattering data the lattice gives an
exact closed map on a finite time window, in which the final state of the medium is fixed by the
data at one point of the spectral plane, and a linear law on the infinite line.

Some questions remain open. The continuum limit is formal and confirmed numerically; a
convergence proof is lacking. The calibration is exact for one soliton parameter only. The
linear law on the infinite line, the discrete analogue of Proposition~2 of
Ref.~\cite{GabitovZakharovMikhailov1985} and the complex extension of the law of Boiti
\textit{et al.}~\cite{BoitiPempinelliPrinariSpire2002}, is derived under the assumption that the
medium returns to the ground state and checked numerically. The return on later rows, with a window that grows with $m$, is observed
but not derived. How the continuum limit of the linear law recovers the continuous evolution,
in which $a$ changes because the medium keeps a polarization after the pulse, remains to be
understood. Natural next steps are the inverted medium and
superfluorescence~\cite{GabitovZakharovMikhailov1984}, a Hamiltonian structure of the lattice in
the MB variables (for the real reduction, Hirota's equation, it is given in
Ref.~\cite{FaddeevVolkov1994}), and the relation of its continuous-time limit to the semi-discrete MB lattices
of Refs.~\cite{Bogoliubov1995,Rybin2001} and to Tsuchida's discretization of the negative
flow~\cite{Tsuchida2015}.

\section*{Data availability}
The data that support the findings of this article are available from the authors upon reasonable request.

\begin{acknowledgments}
We thank A.~Bobenko, S.~Burtsev and I.~Timofeev for valuable discussions and advice.
\end{acknowledgments}

\appendix

\section{Derivation of the lattice equations}
\label{app:derivation}
The continuous $U$ is linear in $\lambda$ and $V$ has a simple pole at $\lambda=0$, and the
$\sigma_2$-symmetry requires coefficient matrices of the form
$\bigl(\begin{smallmatrix}a&b\\-b^*&a^*\end{smallmatrix}\bigr)$. We therefore take
$\mathbf U=\lambda\mathcal A+\mathcal B$ and $\mathbf V=\mathcal F+\mathcal G/\lambda$ with
matrices of this form. A gauge transformation $\Psi_{n,m}\to g_{n,m}\Psi_{n,m}$, with $g_{n,m}$
of the same form, changes $\mathbf U\to g_{n+1,m}\mathbf Ug_{n,m}^{-1}$ and
$\mathbf V\to g_{n,m+1}\mathbf Vg_{n,m}^{-1}$; when $\det\mathcal A\neq0$ and
$\det\mathcal F\neq0$ it brings $\mathcal A$ to $-\ii\sigma_3$ and $\mathcal F$ to $I$; the two conditions are compatible because the $O(\lambda)$ term of~\eqref{eq:compat} gives $\mathcal A_{n,m+1}\mathcal F_{n,m}=\mathcal F_{n+1,m}\mathcal A_{n,m}$. Then
$\mathcal B=(1+C)I+\ii\beta\sigma_3-\tfrac12u$ with real $C$ and $\beta$. The term $\ii\beta\sigma_3$
shifts $\lambda$ in $\mathbf U$, and the zeros of $\det\mathbf U$ are $\lambda=\beta\pm\ii\kappa$;
the requirement that they do not depend on $n$ and $m$ makes $\beta$ a constant. A constant $\beta$
is a detuning: after the shift $\lambda\to\lambda+\beta$, $\mathbf U$ takes the
form~\eqref{eq:Unm} and the pole of $\mathbf V$ moves to $\lambda=-\beta$, which is the detuned
factor of Sec.~\ref{sec:sublayers} with detuning $\epsilon=-\beta$; for a sharp line at resonance we set
$\beta=0$. For $\mathcal G$ we take
the form of the continuous $V$, $\mathcal G=-\hat\rho/(4\ii)$ with traceless Hermitian $\hat\rho$.
This gives~\eqref{eq:Unm} after the lattice steps are absorbed into the variables.

Substituting~\eqref{eq:Unm} into~\eqref{eq:compat}, the terms proportional to $\lambda$ cancel
identically. The terms independent of $\lambda$ give~\eqref{eq:lattice1}
and~\eqref{eq:lattice2}, and the residue at $\lambda=0$,
\[
  \mathbf U_{n,m+1}(0)\,\hat\rho_{n,m}=\hat\rho_{n+1,m}\,\mathbf U_{n,m}(0),
\]
gives~\eqref{eq:lattice3} and~\eqref{eq:lattice4}. Written in real components these are ten
equations for the six real unknowns $E_{n,m+1}$, $C_{n,m+1}$, $\rho_{n+1,m}$, $N_{n+1,m}$; they
are consistent and have the unique solution~\eqref{eq:lattice}. Eliminating $E_{n,m+1}$ and
$C_{n,m+1}$ from the residue condition with~\eqref{eq:lattice1} and~\eqref{eq:lattice2} gives
the explicit form~\eqref{eq:qstep}, with $\det G$ equal to the determinant of the linear system
for the four unknowns. Taking the determinant of~\eqref{eq:compat} and using~\eqref{eq:dets}
and the conservation of $r$ gives $\kappa_{n,m+1}=\kappa_{n,m}$
(Sec.~\ref{sec:normalization}).

\section{The lattice as a two-spin system}
\label{app:tsuchida}
A traceless Hermitian $2\times2$ matrix $S$ with $S^2=k^2I$ can be written as $S=k(2\Pi-I)$ with
a rank-one orthogonal projector $\Pi$, so that
\begin{equation}
  \lambda I+\ii S=(\lambda-\ii k)\Bigl(I+\frac{2\ii k\,\Pi}{\lambda-\ii k}\Bigr).
  \label{eq:projform}
\end{equation}
Applied to~\eqref{eq:spins}, both Lax factors become, up to scalar factors that cancel
in~\eqref{eq:compat}, factors of the form $I+(\alpha_{\rm T}-\beta_{\rm T})P/(\lambda+\beta_{\rm T})$ with
$\beta_{\rm T}=\bar\alpha_{\rm T}$ and a rank-one projector $P$: $\alpha_{\rm T}=\ii\kappa$ for the field factor and
$a_{\rm T}=-\epsilon+\ii r/4$ for the Bloch factor of a class with detuning $\epsilon$. Writing the projectors in
stereographic coordinates, $P\propto(u_{\rm T},1)^T(u_{\rm T},1)^*$ with $u_{\rm T}=-E/[2(\kappa-1-C)]$ for the field
and $q_{\rm T}=\rho/(r-N)$ for the medium, and applying the staggered gauge
$\Psi_{n,m}\to\sigma_3^n\Psi_{n,m}$, which reverses the signs of $E$ and $\rho$ on odd $n$,
Eq.~\eqref{eq:compat} becomes the scalar Hermitian case of the refactorization (3.12) of
Ref.~\cite{Tsuchida2015}; we have checked the identification, including the component equations
of that reference, in exact arithmetic for random data. For real fields, with
$1+C=\cos\alpha$, $E/2=\sin\alpha$, $N=r\cos\Theta$ and $\rho=r\sin\Theta$, the constant gauge
$\mathcal W=e^{-\ii\pi\sigma_1/4}e^{-\ii\pi\sigma_2/4}$ maps $\mathbf U$ and $\mathbf V$ onto the Lax
matrices of Orfanidis~\cite{Orfanidis1978} for Hirota's discrete sine-Gordon
equation~\cite{Hirota1977c}, with spectral parameter $k_{\rm O}=-\lambda$ and $\gamma_{\rm O}=r/4$.

\section{Dressing formulas}
\label{app:dressing}
Let $\lambda_j=\xi_j+\ii\eta_j$, $\eta_j>0$, $j=1,\dots,K$, be distinct points and
$\psi_{j;n,m}=\Psi^{(0)}_{n,m}(\lambda_j)(1,\gamma_j)^T\propto(1,z_{j;n,m})^T$ with
$z_j=\gamma_jq_j^np_j^m$ as in~\eqref{eq:zqp}. The $K$-fold dressing matrix is
\begin{equation}
  \mathcal D(\lambda)=I-\sum_{j,k=1}^K\frac{(M^{-1})_{kj}\,\psi_k\psi_j^\dagger}{\lambda-\bar\lambda_j},
  \qquad M_{jk}=\frac{\psi_j^\dagger\psi_k}{\lambda_k-\bar\lambda_j},
  \label{eq:Kdressing}
\end{equation}
which for $K=1$ reduces to~\eqref{eq:dressing}. With
$\mathcal D=I+\mathcal D_1/\lambda+O(\lambda^{-2})$ the dressed fields are
\begin{equation}
\begin{aligned}
  E_{n,m}&=-2\ii\bigl[(\mathcal D_1)_{12}(n+1,m)+(\mathcal D_1)_{12}(n,m)\bigr],\\
  C_{n,m}&=-\ii\bigl[(\mathcal D_1)_{11}(n+1,m)-(\mathcal D_1)_{11}(n,m)\bigr],
\end{aligned}
  \label{eq:EC_D1}
\end{equation}
and $\hat\rho_{n,m}=\mathcal D_{n,m+1}(0)\,\hat\rho_0\,\mathcal D_{n,m}(0)^{-1}$. In the
normalization $\psi_j=(1,z_j)^T$, $(\mathcal D_1)_{12}=-\Xi$ with
$\Xi=\sum_{jk}(M^{-1})_{kj}\bar z_j$, which gives $E=2\ii(\Xi_{n+1,m}+\Xi_{n,m})$ of
Sec.~\ref{sec:breather}. Since $(\mathcal D_1)_{11}\to-2\ii\sum_j\eta_j$ as $n\to+\infty$ and
$\to0$ as $n\to-\infty$, the second relation in~\eqref{eq:EC_D1} gives
$W=-8\sum_nC_{n,m}=16\sum_j\eta_j$. For the breather, $\lambda_{1,2}=\pm\xi+\ii\eta$ and
$z_2=-\bar z_1=-e^{-X-\ii\Phi}$, one finds
\[
  \Xi=\frac{2\eta\xi\,(\eta\sinh X\cos\Phi+\xi\cosh X\sin\Phi)}{\eta^2\cos^2\Phi+\xi^2\cosh^2X},
\]
which leads to~\eqref{eq:breather}.

\bibliography{refs_discrete-mb_postreview}

\end{document}